\documentclass{aa}  
\usepackage{graphicx}
\usepackage{amsmath}
\usepackage{amssymb}
\usepackage{graphicx}

\usepackage{color}
\usepackage{booktabs}

\usepackage{afterpage}
\usepackage{placeins}
\usepackage{enumitem}

\usepackage{natbib}
\newcommand{\dd}{\mathrm{d}}

\newcommand{\pow}[1]{\ifmmode{}^{#1}\else ${}^{#1}$\fi}
\newcommand{\HI}{{\text{H\MakeUppercase{\romannumeral 1}}}\xspace}

\newcommand{\Lya}{\ifmmode{\mathrm{Ly}\alpha}\else Ly$\alpha$\xspace\fi}

\newcommand{\cm}{\,\ifmmode{{\rm cm}}\else cm\fi}

\newcommand{\ergps}{\,{\rm erg}\,{\rm s}\ifmmode{}^{-1}\else ${}^{-1}$\fi}
\newcommand{\Mpch}{\,{\rm Mpc}\,\ifmmode h^{-1}\else $h^{-1}$\fi}
\newcommand{\snru}{\,\ifmmode{\mathrm{Myr}^{-1}}\else Myr${}^{-1}$\fi}
\newcommand{\kms}{\,\ifmmode{\mathrm{km}\,\mathrm{s}^{-1}}\else km\,s${}^{-1}$\fi}

\newcommand{\onesigma}{$1\hbox{-}\sigma$}

\begin{document}

\title{Modeling $237$ Lyman-$\alpha$ spectra of the ``MUSE-Wide survey''}

\author{Max Gronke
}

   \institute{Institute of Theoretical Astrophysics, University of Oslo, Postboks 1029, 0315 Oslo, Norway\\
              \email{maxbg@astro.uio.no}
             }

   \date{Draft from \today}

  \abstract
{
  We compare $237$ Lyman-$\alpha$ (\Lya) spectra of the ``MUSE-Wide survey'' \citep{Herenz2017} to a suite of radiative transfer simulations consisting of a central luminous source within a concentric, moving shell of neutral gas, and dust. This six parameter shell-model has been used numerously in previous studies, however, on significantly smaller data-sets.
  We find that the shell-model can reproduce the observed spectral shape very well -- better than the also common `Gaussian-minus-Gaussian' model which we also fitted to the dataset.
  Specifically, we find that $\sim 94\%$ of the fits possess a goodness-of-fit value of $p(\chi^2)>0.1$. 
  The large number of spectra allows us to robustly characterize the shell-model parameter range, and consequently, the spectral shapes typical for realistic spectra.
  We find that the vast majority of the \Lya spectral shapes require an outflow and only $\sim 5\%$ are well-fitted through an inflowing shell.
  In addition, we find $\sim 46\%$ of the spectra to be consistent with a neutral hydrogen column density $<10^{17}\cm^{-2}$ -- suggestive of a non-negligible fraction of continuum leakers in the MUSE-Wide sample.
  Furthermore, we correlate the spectral against the \Lya halo properties against each other but do not find any strong correlation.
}


  \maketitle

\section{Introduction}
Not many branches of astrophysics show a similar fast transition from theory- to data-driven as the study of the high-$z$ Universe through Lyman-$\alpha$ (\Lya) emitting objects. While predicted already half a century ago \citep{Partridge1967}, the first detections of high-redshift \Lya emitters was not until more than $25$ years later \citep[e.g.,][]{Moller1993,Rhoads2000}.
They were thus preceded by groundbreaking theoretical studies -- which still represent the solid foundation of our understanding of \Lya radiative transfer in astrophysical environments today \citep{Osterbrock1962,Adams1972,Neufeld1990}.
These theoretical studies were often accompanied by numerical radiative transfer codes \citep[e.g.,][]{Adams1972,Bonilha1979} -- with the focus shifting towards the latter more recently \citep[e.g.,][]{Dijkstra2006,Verhamme2006,Hansen2005}. While this numerical work together with increased computational power allowed for the study of more complex geometries (such as the output from hydrodynamical simulations) as well as the consideration of a wider parameter space, the predictive power has naturally decreased due to this multiplication of degrees of freedom and the associated degeneracies.

In addition to these difficulties, a somewhat unfortunate incident arguably slowed down the progress on theoretical side considerably, namely the apparent non-requirement for (more) complex theoretical models in order to explain a wide range of observed \Lya spectra. Specifically, a homogeneous shell of neutral hydrogen (and dust) surrounding a \Lya and continuum emitting source \citep[as first advocated by ][]{Ahn2002} seems to be able to reproduce many observed line shapes from $z\sim 0$ \citep{Yang2017} to higher redshifts \citep{Verhamme2008,Karman2016}. This has been shown using large libraries of pre-computed line shapes \citep{Schaerer2011,Gronke2015} which can be compared to observations. In spite of this automatizing of the fitting procedure and the consequent common usage of the `shell mode', the physical meaning of its parameters is still unclear.
Also, the structure of the shell-model, that is, the existence of a large solid shell of cold, neutral gas seems somewhat contrived. Nevertheless, the shell-model has been studied numerously and the effect of its parameters on the spectral shape are understood well \citep{Verhamme2014}. These facts make the shell-model a common starting point of theoretical modeling of \Lya propagation -- and it does so since more than a decade.

This rather slow progress
on the theoretical side has not halted the planning and execution of new experiments geared towards \Lya detection and study. As a result, the pure number of detected \Lya emitting objects -- even at the highest observable redshifts -- has increased impressively (e.g., the latest detection of $\sim 2000$ \Lya emitters at $z\sim 6-7$ \citealp{2017arXiv170407455O}; also see \citealp{Barnes2014} for a list of \Lya surveys), and the quality of the detections has improved significantly \citep[e.g.,][]{2017arXiv170609428M}. This development has not only lead to the discovery of new types of astrophysical objects such as `Lyman-$\alpha$ blobs' \citep[e.g.,][]{1996ApJ...457..490F,1999MNRAS.305..849F,Steidel2000,North2017}, `giant \Lya halos' \citep[e.g.,][]{Cantalupo2014,Hennawi2015,Cai2016}, and `extreme equivalent width objects' \citep[e.g.,][]{Sobral2015} but also opened up the chance of statistical usage of \Lya data, for example, to constrain the `Epoch of Reionization' \citep[see review by][]{Dijkstra2014a} or the leakage of ionizing photons \citep[e.g.,][]{Dijkstra2016b,Verhamme2016a}.

The latest stage in this series of technological advances and consequent observational progress has been the implementation of ESO's `Multi Unit Spectroscopic Explorer' (MUSE; \citealp{2010SPIE.7735E..08B}). Equipped with $24$ integral field units, this integral field spectrograph has literally added an additional dimension to \Lya observations and will thus multiply the available \Lya data even further. The unprecedented sensitivity of MUSE has already revealed the ubiquity of \Lya emission throughout the Universe \citep{2015A&A...575A..75B,Wisotzki2015,2017MNRAS.471..267D}, and the combination of imaging with spectral information opens up many new possibilities for \Lya data analysis -- which will further our understanding of astrophysical \Lya emitting objects.

Recently, the data of the first installment of the MUSE-Wide survey has been released publicly \citep{Herenz2017} which contains $237$ \Lya spectra around $z\sim 4$. The sample (corresponding to $\sim 20\%$ of the final survey area) is emission line-selected which allows an unbiased spectral analysis.
We take this opportunity to assemble a statistical relevant sample of shell-model fits. Thus far, the number of modeled spectra found in the literature are often merely individual spectra \citep[e.g.][]{2016ApJ...821L..27V,Dahle2016} or compilations up to $\sim 10$s \citep{Hashimoto2015,Karman2016,Yang2017}. A (much) greater number of spectra will, thus, allow us to address the following questions:
\begin{itemize}
\item How well can the shell-model reproduce observed spectral line shapes? Is there, e.g., a certain spectral characteristic that cannot be reproduced? 
\item What are common values for shell-model parameters? Answering this question is useful for studies concerned with a single object and might help them to characterize how unusual their spectral shape is.
\item Do any correlations amongst the shell-model- or with external parameters exist?
\end{itemize}
Furthermore, this work will provide the foundational data for future work `decrypting' the shell-model parameters. In a recent series of papers \citep{Gronke2016a,Gronke2016b,2017arXiv170406278G}, we've shown already that the shell-model parameters do not reflect the physical properties of a simple multiphase model. However, in such a case the spectra do commonly also show more flux a line-center and are less asymmetric (in spite of outflows) compared to observed line-profiles. However, in the presence of small-scale structure such as tiny `cloudlets' \citep[as theoretically predicted by][]{McCourt2016}, \Lya photons escape the -- still multiphase -- system as it was homogeneous which yields more realistic spectra, and makes the obtained shell-model parameters physically meaningful. In this work, we will interpret the data also in the light of these recent results.

The paper is structured as follows: Sec. \ref{sec:method} describes the MUSE-Widefield data set, and the fitting procedure. We present and discuss our results in Sec.~\ref{sec:results}, and conclude in Sec.~\ref{sec:conclusion}.
When required, we use the \citet{2016A&A...594A..13P} parameters of a flat $\Lambda$CDM Universe ($H_0 = 67.7\,\kms\,\mathrm{Mpc}^{-1}$, $\Omega_{\mathrm{m0}}=0.307$, $\Omega_{\Lambda0}=0.691$), and use `$\log$' as logarithm with base ten.

\section{Method}
\label{sec:method}

\subsection{The MUSE-Wide data}
\label{sec:MW-data}

The MUSE-Wide catalogue used in this work was described extensively in \citet{Herenz2017}. Here, we summarize the main properties of the data set and explain which pre-processing steps we undertook prior to modeling.

In total, the catalogue contains $831$ datasets with a spectral resolution of $2.5\,$\AA\ out of which $237$ contain a \Lya line. These \Lya emitting objects are in a redshift range of $z_{\mathrm{MW}} = [2.97,\,6.00]$ with a median redshift of $3.83$. The redshift $z_{\mathrm{MW}}$ given for these objects stems from fitting an asymmetric profile to the \Lya line \citep[see \S~3.4.1 in][]{Herenz2017} in order to obtain the peak wavelength which was converted to a redshift. The quoted redshift error is due to the fitting procedure, i.e., corresponds to an uncertainty in the peak position. This means for both, the redshift and the redshift error, no correction for radiative transfer effects was introduced. During the fitting procedure, we will leave the redshift as a free parameter. In order to use the spectra, we extract a wavelength range which allows for a shift of $\Delta z = [-0.02,\,0.01]$ from the given peak-redshift while considering that even with a maximum shift no data point is outside the range of $[\pm 2500\,\kms]$ which is supported by the fitting pipeline. The asymmetric $\Delta z$ range is due to the fact that most \Lya peaks are redwards of non-resonant lines, i.e., we expect in most cases a $z$ value blueward of $z_{\mathrm{MW}}$  (i.e., $z < z_{\mathrm{MW}}$). However, we still leave enough room ($\Delta z=0.01$ corresponds to $\sim 750\kms$ at $z=3$ -- a value larger than commonly found shifts of several $\sim 100\kms$ \citealp{Shapley2003,Steidel2010a,Kulas2011}) for an opposite shift. This procedure ensures that even while shifted the same number of data points is compared to the theoretical spectrum.

As some spectral data contain negative mean continuum fluxes -- which are unphysical -- we take a conservative approach, and remove this continuum level from all the spectra. This implies, that we will only use the \Lya line shape to fit theoretical spectra to it and ignore the surrounding continuum.

Another important property of the \citet{Herenz2017} dataset used in this work include the `confidence' category which can be either $3$ for spectra which are almost certainly \Lya ($\sim 40$\% of the spectra), $2$ for near-certainty ($\sim 52$\%), and $1$ for unsure detections which are, however, not spurious ($\sim 8\,$\%). We refer to \citet{Herenz2017} for more details about the classification.

\begin{figure}
  \centering
  \includegraphics[width=.95\linewidth]{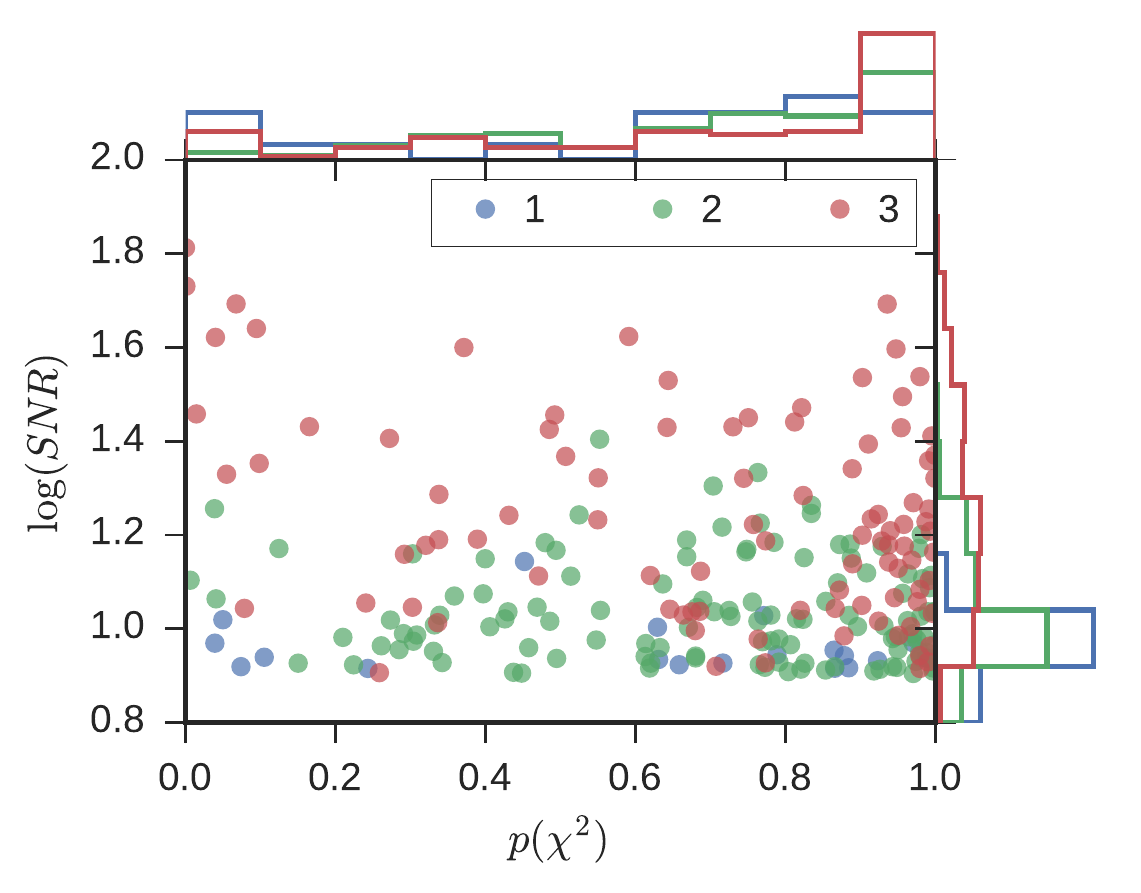}
  \caption{The signal-to-noise ratio of the \Lya line ($SNR$) versus the quality of the fit ($p(\chi^2)$ computed as described in \S~\ref{sec:fit-quality}). The color coding denotes the `confidence' category as defined in \citet{Herenz2017}.}
  \label{fig:chisq_vs_snr}
\end{figure}

\begin{figure*}
  \centering
  \includegraphics[width=\textwidth]{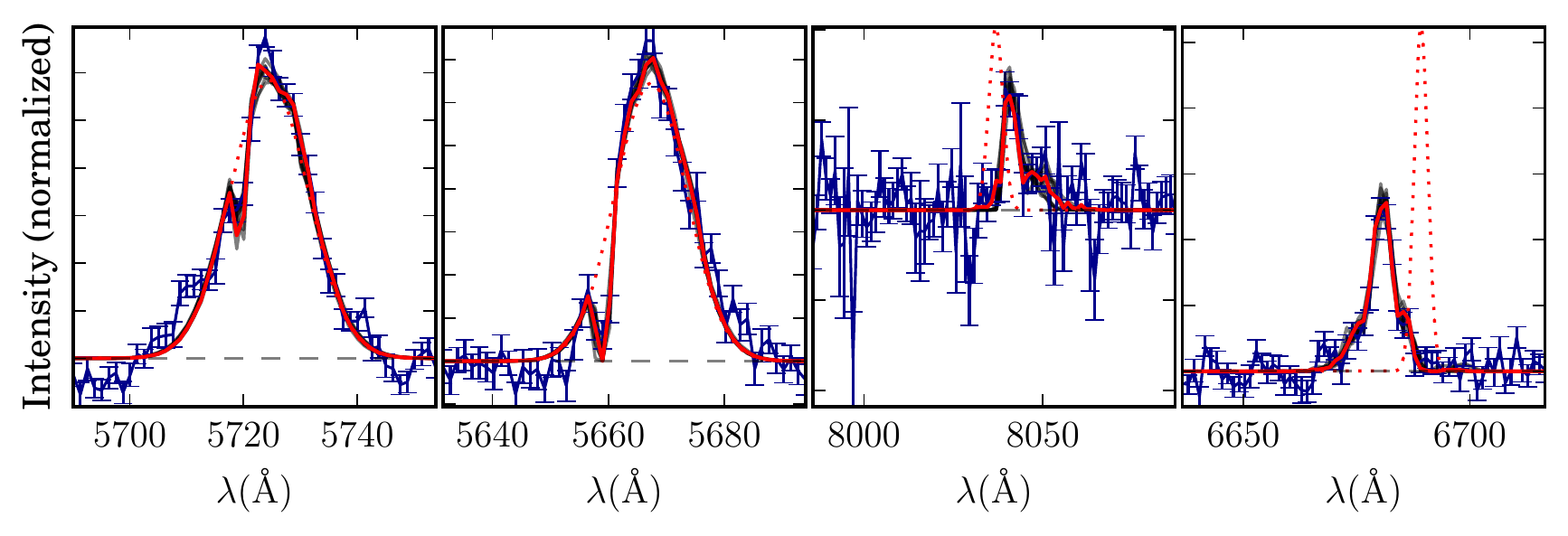}
  \vspace{-.5cm}
  \caption{The four spectra with the lowest $p(\chi^2)$-values of (from left- to rightmost panel) $\{<0.1,\,0.5, 6, 14\}\times 10^{-3}$. The data is shown in dark-blue, and the best fit shell-model spectrum with a solid red line (the dotted red line is the intrinsic spectrum). In addition, the semi-transparent black lines show $10$ randomly drawn spectra from the burned-in chains. Apart from the spectrum in the third panel which has been graded with `confidence' level $2$ the other three spectra have been classified as `3'.}
  \label{fig:spectra_worstfit}
\end{figure*}

\subsection{Shell-model fitting}
\label{sec:sm-fitting}
The shell-model fits were carried out via an improved version of the fitting procedure described in \citet{Gronke2015}. Because the shell-model consists of a central source emitting \Lya photons inside a moving shell of neutral hydrogen and dust, it can be parametrized through the \HI column density $N_{\HI}$, the (all-absorbing) dust optical depth $\tau_{\mathrm{d}}$, the outflow velocity $v_{\mathrm{exp}}$ ($<0$ for infall), the intrinsic width of the \Lya line $\sigma_{\mathrm{i}}$, and the effective temperature of the gas $T$.

In order to fit observed spectra by this model, one usually divides these parameters into sets of discrete and continuous ones. While the parameter space spanned by the discrete ones has to be covered through (computationally relatively expensive) radiative transfer simulations, the continuous ones can be modeled in post-processing. We model the effect of $\sigma_{\mathrm{i}}$ and $\tau_{\mathrm{d}}$ through individual weighting of photon packages \citep[see][for details]{Gronke2015} with an allowed range of $\sigma_{\mathrm{i}} \in [1,\,800]\kms$ and $\tau_{\mathrm{d}}\in [0,\, 5]$. The remaining subspace spanned by $(v_{\mathrm{exp}},\,\log N_\HI / {\mathrm{cm^{-2}}},\,\log T / {\mathrm{K}})$ we cover via a $(54,\,30,\,8)$-grid which totals $12960$ discrete models. Each of the models was carried out using $20,000$ `\Lya' photons with Gaussian intrinsic spectrum centered at $0$ and a standard deviation of $800\kms$ and $150,000$ `UV' photons with uniform intrinsic spectrum in the range $\pm 2700\kms$.

While the parameters $\log N_\HI / {\mathrm{cm^{-2}}}$ and $\log T / {\mathrm{K}}$ are spaced uniformly between their respective limits $[16,\,21.8]$ and $[3,\,5.8]$, the outflow velocity $v_{\mathrm{exp}}$ can take the additional values of $2\kms$, $5\kms$, and $8\kms$ apart from the values with $10\kms$ separation between $0$ and $490\kms$. This is due to the strong effects small velocities can have on the resulting spectrum due to the peaked nature of the \Lya scattering cross-section. Furthermore, we use the symmetry of \Lya radiative transfer to cover inflow velocities \citep{2017arXiv170403416D}.

In summary, the following steps were carried out for each spectrum:
\begin{enumerate}[itemsep=10pt]
\item We fix a redshift prior of with mean $z_{\mathrm{MW}} - 0.005$ and standard deviation of $\sigma_z = 0.005$. As stated above, the offset is due to the more likely observation of the red peak of the \Lya line. This prior is very wide and easily overruled by the data.
\item Due to the multi-modal nature of the likelihood landscape -- with many local peaks which are, however, of dramatically different altitude -- it is crucial to find the global likelihood maximum around which one can carry out the Monte-Carlo optimization. In a continuous parameter space with unknown likelihood function it is impossible to know whether or not this global maximization is successful. However, we do our best by combining a basinhopping algorithm over the discrete parameters with a local maximization over the continuous parameters. This means, we perform a basinhopping algorithm with randomly chosen starting position until no better results is obtained\footnote{In practice, between $\sim 120$ and $350$ are performed.}. Each step consists of a random displacement in parameter space followed by a local maximization using the robust \texttt{POWELL} algorithm. In order to speed up the global maximization process by reducing the memory access, we execute the basinhopping over a local maximization using only the continuous parameters ($z$, $\tau_{\mathrm{d}}$, $\sigma_{\mathrm{i}}$). For this process, we use the \texttt{COBYLA} method \citep{powell1994direct}.
  
\item From this initial position, we start the affine invariant Monte-Carlo sampler \texttt{emcee} \citep{Goodman2010,Foreman-Mackey2012} with $200$ walkers and $1240$ steps. Specifically, we place the walkers on several positions identified through the optimization process weighted by their likelihood. In case a better-fitting spectrum ($\Delta\ln p \ge 0.5$) than the starting one is found, the MC sampling is restarted from this position (with $1200$ steps).
  
\item For further analysis, we discard the first `burn-in' steps for which the mean (over the walkers) of each parameter does not disagree by more than $15\%$ off the final value. From the remaining chains we compute the median values of each shell-model parameter and its uncertainties.
\end{enumerate}

\begin{figure*}
  \centering
  \includegraphics[width=\textwidth]{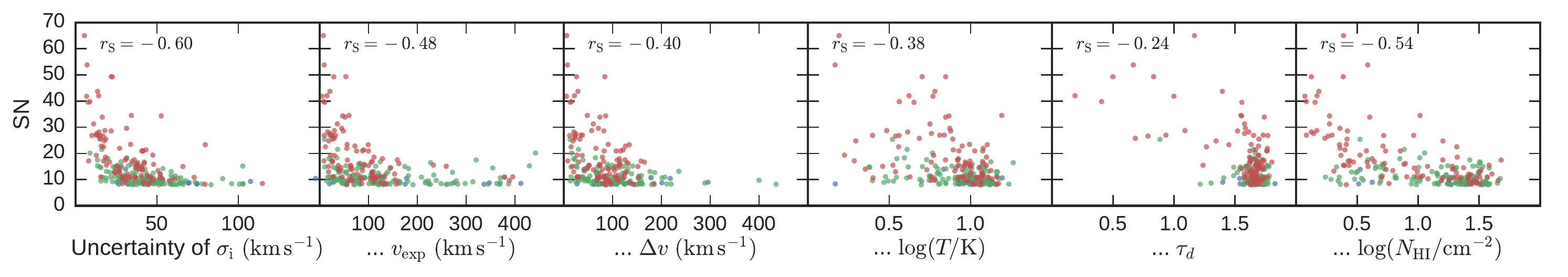}
  \caption{Uncertainties on the obtained fitting parameters as a function of the signal-to-noise ratio. Specifically, we show half of the $84$th minus the $16$th percentile of the corresponding posterior distributions. The color corresponds to the `confidence' category assigned to each spectrum by \citet{Herenz2017} with \textrm{red} for the highest confidence (3), followed by \textrm{green} and \textrm{blue} for the lowest confidence class.} 
  \label{fig:sm_error}
\end{figure*}

\section{Results}
\label{sec:results}

\subsection{Fit quality}
\label{sec:fit-quality}
In order to assess the quality of the shell-model fits, we computed the $p(\chi^2)$ values of the best-fit spectra.
Following \citet{pearson1900x} we compute $\chi^2 = \sum_i (y_i^{\mathrm{(data)}} - y_i^{(\mathrm{fit})})^2 / \sigma_i^2$ (where $\sigma_i$ is the flux error in bin $i$ and the sum is taken over the $N$ data points of the spectrum), and quote as $p(\chi^2)$ the cumulative probability of a $\chi^2$-distribution with $N$ degrees of freedom having values greater than $\chi^2$.

Figure~\ref{fig:chisq_vs_snr} shows the distribution of the resulting $p(\chi^2)$ values as a function of the signal-to-noise ratio of the \Lya line. Also shown in Fig.~\ref{fig:chisq_vs_snr} is the ``confidence'' class of the spectrum (as discussed in \S~\ref{sec:MW-data}). One can notice that the $p(\chi^2)$ values are roughly uniformly distributed between the extremes of $3\times 10^{-13}$ and $0.994$ with a median of $0.773$. Overall, the fit quality is very good with $94\%$ [$73\%$] of the spectra with a $p(\chi^2)>0.1$ [$>0.5$].

As visible from Fig.~\ref{fig:chisq_vs_snr}, there seems to be no strong overall correlation neither versus the $SNR$ (Spearman's rank correlation coefficient of $r_{\mathrm{S}}\approx-0.08$, with a two-sided $p$-value of a hypothesis test to check whether the two variables are correlated $p_{\mathrm{S}}\approx 0.24$) nor with any of the (median) shell-model parameters ($p_{\mathrm{S}} >10^{-3}$)-- with the exception of $\sigma_{\mathrm{i}}$ where $r_{\mathrm{S}}\approx -0.26$ ($p_{\mathrm{S}}\approx 4\times 10^{-5}$). This tentative anti-correlation between $\sigma_{\mathrm{i}}$ and the shell-model fit quality might be suggestive of a stronger influence of radiative-transfer for spectra with lower $\sigma_{\mathrm{i}}$.

Figure~\ref{fig:spectra_worstfit} shows the four worst fits, i.e., with the lowest $p(\chi^2)$ values. It is noteworthy that even for those spectra the fit does not ``look that bad'', i.e., the model does not seem to fail catastrophically. The large $\chi^2$ values in the examples shown in Fig.~\ref{fig:spectra_worstfit} can be attributed to slightly lower (negative) continuum on the blue side of the emission (two leftmost panels), or some (again with negative intensity) outliers (two rightmost panels). When increasing the error bars only by $10\%$ ($20\%$), the overall $p(\chi^2)$ increases from $0.77^{+0.20}_{-0.44}$ to $0.95^{+0.05}_{-0.22}$ ($0.99^{+0.01}_{-0.06}$).\footnote{Here -- the same as the rest of this work -- we use the notation $A^{b}_{c}$ where $A$ denotes the median and $b$ ($c$) the difference to the $16$th and $84$th percentile.} This would not be the case if the theoretical model cannot reproduce certain features which are present in the observed spectra -- such as an additional peak -- and lets us conclude that overall, the shell-model seems to provide (surprisingly) good fits to all the $237$ \Lya spectra.

\begin{figure*}
  \centering
  \includegraphics[width=.95\textwidth]{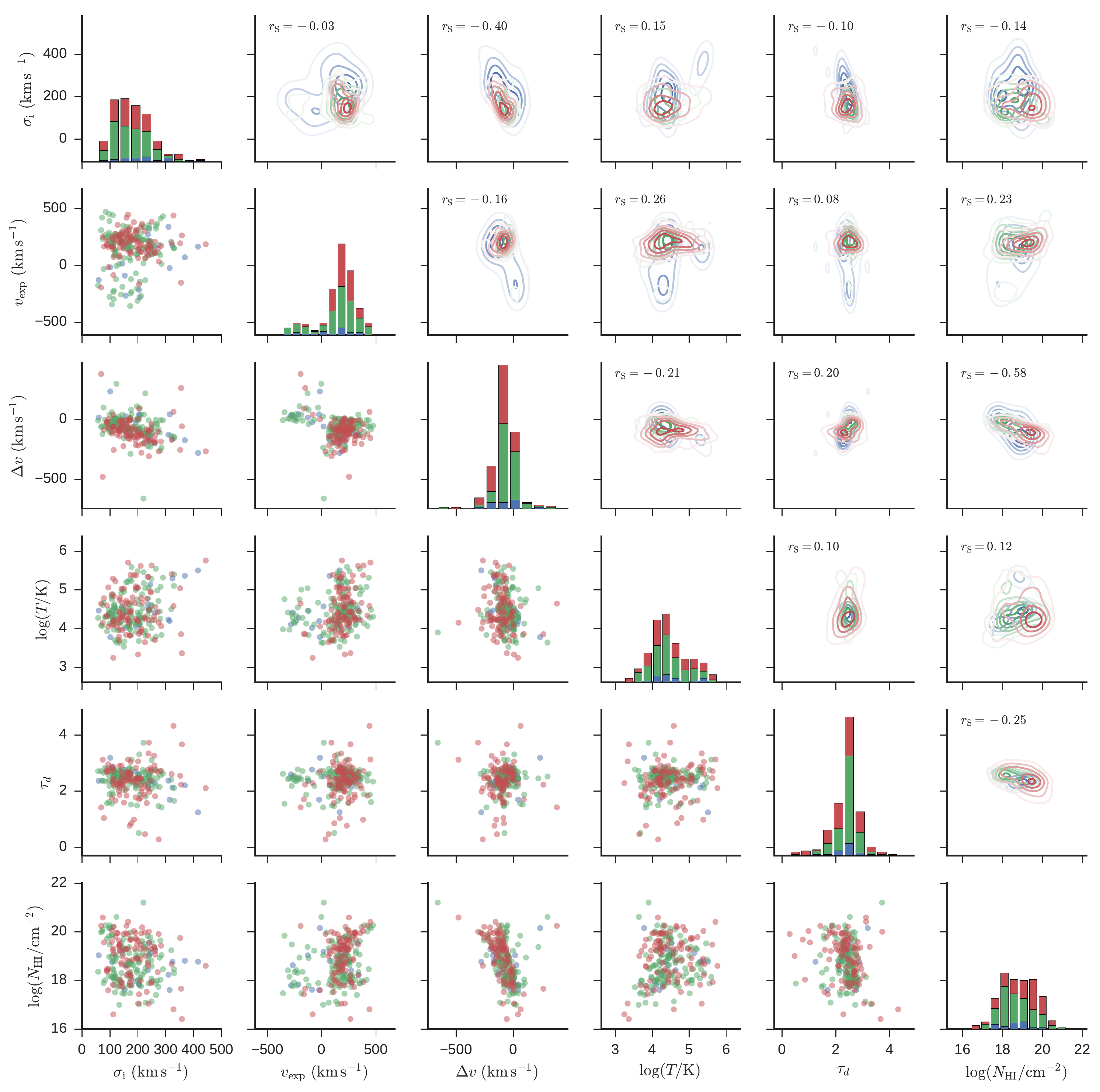}
  \caption{Correlation matrix between the median obtained shell-model parameters. The color coding denotes the `confidence' category the spectrum was assigned (blue $= 1$, green $= 2$, red $ = 3$). While the lower triangle shows scatter plots with each spectrum marked individually, the same data is plotted as contours of kernel-density estimators in the upper right triangle. The panels along the diagonal show \textit{stacked} histograms of the (projected) distribution.}
  \label{fig:shell-model-triangle}
\end{figure*}

\begin{figure*}
  \centering
  \includegraphics[width=\textwidth]{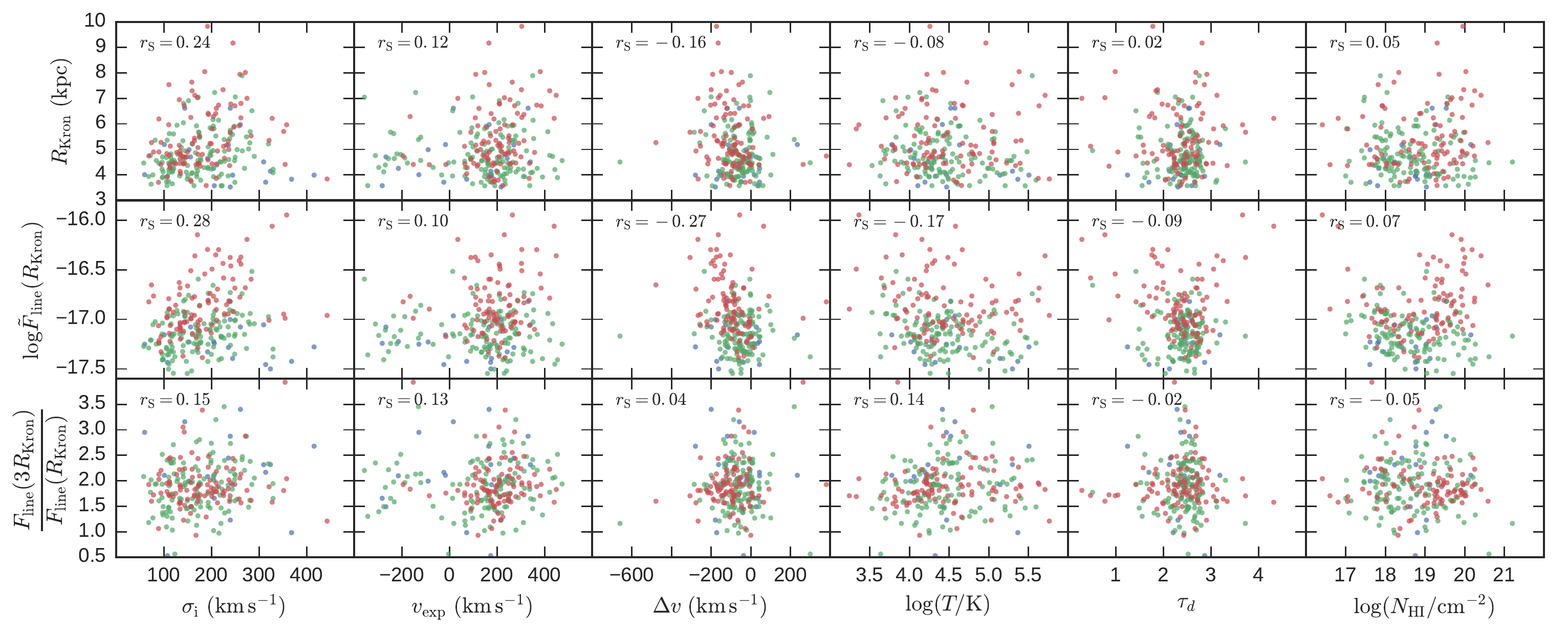}
  \caption{The extent of the \Lya halo and it's brightness versus the six obtained shell-model parameters. The upper and central rows show the Kron radius (in kpc) and the line flux (with $\tilde F = F / (\ergps\,\cm^{-2})$) as a function of the fitting parameters (see \S~\ref{sec:haloprops} for more details). The lower row shows the flux ratio between $3\times$ and $1\times R_{\mathrm{Kron}}$. The color coding corresponds again to the `confidence' category (blue, green, red in increasing order). Also shown in each panel are the Spearman correlation coefficients.}
  \label{fig:haloparams}
\end{figure*}

\subsection{Fitting results}
\label{sec:fitting-results}

Through the Monte-Carlo sampling, we obtain an estimate of the full posterior distribution which we quantify through one- and two-dimensional projections, and the $16$th, $50$th and $84$th percentiles. Due to the large number of data points, we cannot show all the three percentiles (i.e., the median and the uncertainty on the parameters) in the same plot. Therefore, Fig.~\ref{fig:sm_error} shows first the uncertainty on the fitting parameters which we quantified through $\frac{1}{2}(q_{84} - q_{16})$ (where $q_X$ it the $X$th percentile) as a function of the signal-to-noise ratio. Also, the color coding shows the `confidence' group with red the highest  and blue the lowest confidence spectra. Clearly, with improved spectral quality, the fitting uncertainty reduces. For instance, in order to obtain a $\lesssim 0.5\,$dex uncertainty on the column density, the signal-to-noise ratio should be $\gtrsim 20$. However, as is also apparent by comparing the various panels of Fig.~\ref{fig:sm_error} the improvement is not equally strong for the six parameters. For instance, the dust-optical depth shows mostly (with a few exceptions) a very large uncertainty ($\gtrsim 1$), and the effective temperature can only be estimated within an order of magnitude \citep[see also][for similar results using mock spectra]{Gronke2015}.

Note that a while a greater signal-to-noise ratio will improve the constraints on the shell-model parameters, the increased spectral resolution will -- most likely -- not help significantly. The MUSE spectral resolution of $2.5\,$\AA\ corresponds to a resolving power of $R\sim [1900,\,3400]$ for the \Lya line in the considered redshift-range. However, above $R\sim 1000$ the uncertainty does not depend significantly on the spectral resolution  \citep[see Appendix A in][]{Gronke2015}. Better resolution spectra provide, nevertheless, important additional information as potentially features arise which cannot be captured by the shell-model (as, e.g., the extended red-wing in \citealp{2017arXiv170609428M} or the reversed absorption draught in \citealp{2016ApJ...820..130Y}) let us peek beyond its simplicity. They might, thus, give insights into the physical meaning of the shell-model parameters\footnote{Naturally, \textit{if} such spectral features are present they would lower the fit-quality of the shell-model discussed in \S~\ref{sec:fit-quality}. However, this has to be tested with higher-resolution data as the current data-set seems to be well fit by the shell-model.}.
\\

Figure~\ref{fig:shell-model-triangle} shows the one- and two-dimensional projections of the distribution of the obtained median shell-model parameters. Here, we again split the data-set according to the `confidence' group provided. No strong correlation is visible between the parameters. A slight tendency for more shift (which we show in velocity space as $\Delta v$; see below) for larger \HI column densities or larger $\sigma_{\mathrm{i}}$ are the only (weak) correlations we could identify.

The distribution of each individual parameter might help to understand what the shell-model parameters mean physically. Here we discuss them one-by-one:
\begin{itemize}[itemsep=10pt]
\item the neutral hydrogen column density is near log-normally distributed (\citealp{d1973tests} test of normalization yields $p_{\mathrm{AP}}\approx 0.14$) with mean $18.78$ and standard deviation $0.88$. This means the limits on this parameter are sufficiently far away, i.e., a wider covering of the parameter space would most probable not help to obtain better fits.
Interestingly, $\sim 46\%$ of the spectral shapes [$\sim 36\%$ of the `category $3$' spectra] are consistent (within \onesigma) with a column density of $N_\HI < 10^{17}\cm^{-2}$. This could indicate a potentially large number of Lyman-continuum leakers amongst the MUSE-Wide galaxies. However, better quality spectra are required (in particular, for the fainter galaxies) in order to draw any conclusions.  With the current data set only one object clearly prefers such a low column density, that is, the $84$th percentile is lower than the threshold.
We would also like to caution that even if this suspicion substantiates with better quality data, i.e., also then the \Lya spectrum points towards $N_{\HI}<10^{17}\cm^{-2}$, it is not clear if these objects would be detected as continuum leakers due to a potential (non-)directional dependence of the \Lya spectrum (\citealp{2016ApJ...823...74D}; Eide et al., in prep.; but see \citealp{2012A&A...546A.111V}).

\item the `effective temperature' of the shell obtains most likely takes values of $T\sim 10^{4}\,$K (median $\log(T / \mathrm{K} )\sim 4.38$) which is close to what is expected of neutral gas. However, also larger (and smaller) values are found -- and, in particular temperatures of $T\sim 10^5\,$K which are thermally unstable. If the shell-model temperature had some physical meaning, these values could be explained with subgrid turbulence which enters as $T_{\mathrm{eff}} = T_0 + \langle v^2 \rangle m / k_{\mathrm{B}}$, i.e., dispersion velocities of $\sqrt{\langle v^2  \rangle}\sim 25\kms$ would correspond to an effective temperature of $T\sim 10^5\,$K. Note also that mostly $\sim 1\,$dex uncertainty is associated with the effective temperature (cf Fig.~\ref{fig:sm_error}). 
  
\item the width of the intrinsic spectrum $\sigma_{\mathrm{i}}$ occupies the range $\sim 172^{+75}_{-60}\kms$. The distribution is furthermore faster declining towards smaller values of $\sigma_{\mathrm{i}}$ which makes the canonical value of $\sigma_{\mathit{i}}\sim 13\kms$ -- corresponding to thermal motion of a gas with $T\sim 10^4\,$K -- unlikely. However, both turbulent motion of the emitting gas (in the case of, e.g., \Lya production through fluorescence \citealp{Hogan1987,Mas-Ribas2016a}) and of cold gas relatively nearby the emitting source will lead to a broadening of the \Lya line prior to additional radiative transfer effects, and these larger $\sigma_{\mathrm{i}}$ values are relatively easy to justify. Interestingly, although even broader intrinsic lines of $\sigma_{\mathrm{i}} \gtrsim 400\kms$ are allowed by the fitting pipeline (up to $\sigma_{\mathrm{i}} \le  800\kms$), they are not favored by the data.
  
\item the resulting (absorbing) dust optical depth of the shell-model is centered around $\tau_{\mathrm{d}}\sim 2.45^{+0.29}_{-0.41}$. Within the geometry of the homogeneous shell, these values would correspond to extremely low escape fractions of $\sim 0.037^{+0.028}_{-0.016}$ which seem highly unlikely for the considered objects.
Furthermore, if converted to a metallicity (using the SMC conversion factor of $Z/Z_{\odot} = 9.3\times 10^{20} \cm^{-2}/N_{\HI}\tau_{\mathrm{d}}$ which assumes an albedo of $A=0.32$ \citealp{Pei1992,2001ApJ...554..778L,Laursen2009}) the resulting metallicities of $\log (Z/Z_{\odot})=2.5^{+0.93}_{-0.92}$ are unrealistically high.

Therefore, the literal interpretation of the dust optical depth is not very probable. 
However, as the uncertainty on the $\tau_{\mathrm{d}}$ is mostly greater than unity (compare Fig.~\ref{fig:sm_error}), lower dust contents are consistent with the spectral shape.
The reason for these preferred large $\tau_{\mathrm{d}}$ values can be understood when considering of how an increased dust optical depth changes the emergent \Lya spectrum. A common misconception -- probably due to the (near-)frequency independence of the dust cross-section around $\lambda\approx 1216\,$\AA\ \citep{Pei1992,Laursen2009} -- is that the effect of dust does not change the \Lya spectrum (but merely the \Lya escape fraction). While this assumption holds for many cases, it does not universally. This is because dust attenuates the photons proportional to their path length through cold, dusty medium, and thus, if there is a correlation between emergent frequency and trajectory length, the shape of the spectrum will be changed. For instance, in `usual' double-peaked \Lya profiles through outflowing media (with an enhanced red side), the \Lya photons forming the blue peak have travelled further through neutral hydrogen. Therefore, an increased dust content will lower the blue peak and increase the spectral asymmetry. This might be the reason for the high $\tau_{\mathrm{d}}$ preference of the data. However, for the observed spectra the blue side might be attenuated not by dust but by neutral gas in the CGM and / or IGM which scatters \Lya photons out of the line-of-sight, and thus, effectively absorb the blue side of the spectrum \citep{Dijkstra2007a,Laursen2011}.
  
\item the shell outflow velocities (i.e., $v_{\mathrm{exp}} > 0$) are approximately normally distributed with mean $\sim 211\kms$ and standard deviation $\sim 94\kms$. For lower quality spectra (categories $1$ \& $2$), we found also inflow velocities with similar magnitude. The values of $v_{\mathrm{exp}}$ thus agree with previous measured outflow velocities in the nearby \citep{1998A&A...334...11K,Rivera-thorsen2014} and distant Universe \citep{Shapley2003,Kulas2011}. Overall, we find (within \onesigma) only $\sim 5\%$ of the spectra to be better fit by inflow velocities [$\sim 3\%$ of the confidence $3$ spectra] and $\sim 25\%$ [$\sim 9\%$] to be consistent with inflows. These limits could potentially constrain the covering fraction of cold, inflowing gas and thus yield insights into the gas budget of high-$z$ galaxies. We caution, however, that the directional dependence of \Lya spectra (in such scenarios) is yet unclear \citep[see figure D.1 in][for an angle-averaged spectrum of an filamentary accreting object which shows signs of inflows]{Trebitsch2016}.

\item while not technically being a shell-model parameter, the redshift of the source is crucial for the fit. We display the redshift in Fig.~\ref{fig:shell-model-triangle} as shift between the \Lya peak $z_{\mathrm{MW}} $\citep[as provided by][]{Herenz2017} and the modeled intrinsic redshift $z$. Furthermore, we convert the difference to velocity space resulting in
  \begin{equation}
    \label{eq:Deltav}
    \Delta v = c\frac{z - z_{\mathrm{MW}}}{1 + z_{\mathrm{MW}}}\;.
  \end{equation}
  We find $\Delta v$ values of $\sim -76_{+77}^{-79}\kms$ ($\sim -95_{+78}^{-96}\kms$ for only the highest `confidence' spectra), which implies an emission blueward of the main spectral peak for the majority of analyzed spectra -- as commonly found for other \Lya emitting objects \citep[e.g.,][]{Erb2014,Trainor2015}. Only $\sim 3.8$\% ($\sim 4.2\%$ of the `confidence' $3$ spectra) of the spectra show $\Delta v > 0$ (above \onesigma). On the other hand, $\sim 49\%$ of the spectra ($\sim 29\%$ of the `confidence' three spectra) are consistent with $\Delta z \sim 0$ (within \onesigma).
  Another potentially interesting point is the correlation between $N_{\HI}$ and $\Delta z$, i.e., larger shifts for higher column densities which follows directly from the frequency diffusion in the models. Observationally, it has been established that this shift is anti-correlated with the \Lya equivalent width \citep{Erb2014} which invites to the interpretation of facilitated \Lya escape in systems with less frequency diffusion.
\end{itemize}

Note that we did not find an evolution of any of the above discussed parameters with redshift. We found, however, that only spectra with high SNR ($\gtrsim 20$) occupy the parameter space with $\Delta v \lesssim 200\kms$ or $N_{\HI} \gtrsim 10^{19.5}\cm^{-2}$ -- similarly to what we discussed above for $v_{\mathrm{exp}}$. This is likely due to the fact that the spectral features required for, e.g., a large column density (for instance, the extended wing) is only apparent for higher quality spectra.\\

All the above described shell-model parameters, that is, the $16$th, $50$th, and $84$th percentiles of the posterior distributions for each object are publicly available\footnote{The data can be downloaded at \url{http://bit.ly/a-spectra-of-MW}.}.

\subsection{Correlations with halo properties}
\label{sec:haloprops}
The combined modeling of both the spectrum and surface brightness distribution is a crucial next step to understand how \Lya photons escape galaxies and what information they obtain on their trajectory as using both constraints breaks degeneracies which otherwise exist. Integral field spectrograph such as MUSE are the perfect tool to study these two angles simultaneously. As a first step towards the full joined modeling, we want to correlate the spectral with spatial information.

Included in the MUSE-Wide catalogue is the \citet{1980ApJS...43..305K} radius $R_{\mathrm{Kron}}$ of each object as calculated by \texttt{LSDcat} \citep{Herenz2017a} as well as the flux within $\{1,2,3\}\times R_{\mathrm{Kron}}$. \texttt{LSDcat} calculates $R_{\mathrm{Kron}}$ following the two-dimensional definition of \citet{1996A&AS..117..393B} which reads
\begin{equation}
  R_{\mathrm{Kron}} = \frac{\sum r I(r)}{\sum I(r)}
  \label{eq:kron2D}
\end{equation}
where the summations are carried out over the $2$D pseudo-narrowband images created from the $3$D \textrm{MUSE} datacubes.

Figure~\ref{fig:haloparams} shows the $R_{\mathrm{Kron}}$ (upper row), the flux within  $R_{\mathrm{Kron}}$ (central row), and the ratio of the flux within $3 R_{\mathrm{Kron}}$ and $R_{\mathrm{Kron}}$ (lower row) versus the six shell-model parameters. The latter quantity is a measure how steep the surface brightness profile is falling off. In each panel, we also display the Spearman correlation coefficient. We cannot identify a clear correlation between the   fitting parameters and the halo properties. However, some tentative correlations might be apparent. For instance, $\sigma_{\mathrm{i}}$ and the halo extent as well as the brightness of the source -- in particular the high quality spectra (shown as red points) are somewhat correlated. Also $\Delta v$ and the flux within $R_{\mathrm{Kron}}$ show some level of anti-correlation. In order to confirm or rule out these tentative correlations (as well as the ones discussed in \S~\ref{sec:fitting-results}), the uncertainty of the shell-model parameters should be lower than the correlation range of interest. This mean a similar number of spectra with better signal-to-noise ($SNR\gtrsim 20$; cf. Fig.~\ref{fig:sm_error}) would be most useful -- a goal which can be achieved in the near future given that this first release only covers about one fifth of the total survey area, and the ``MUSE-Wide'' scanning strategy is very shallow \citep[one hour per field][]{Herenz2017}.

\begin{figure}
  \centering
  \includegraphics[width=\linewidth]{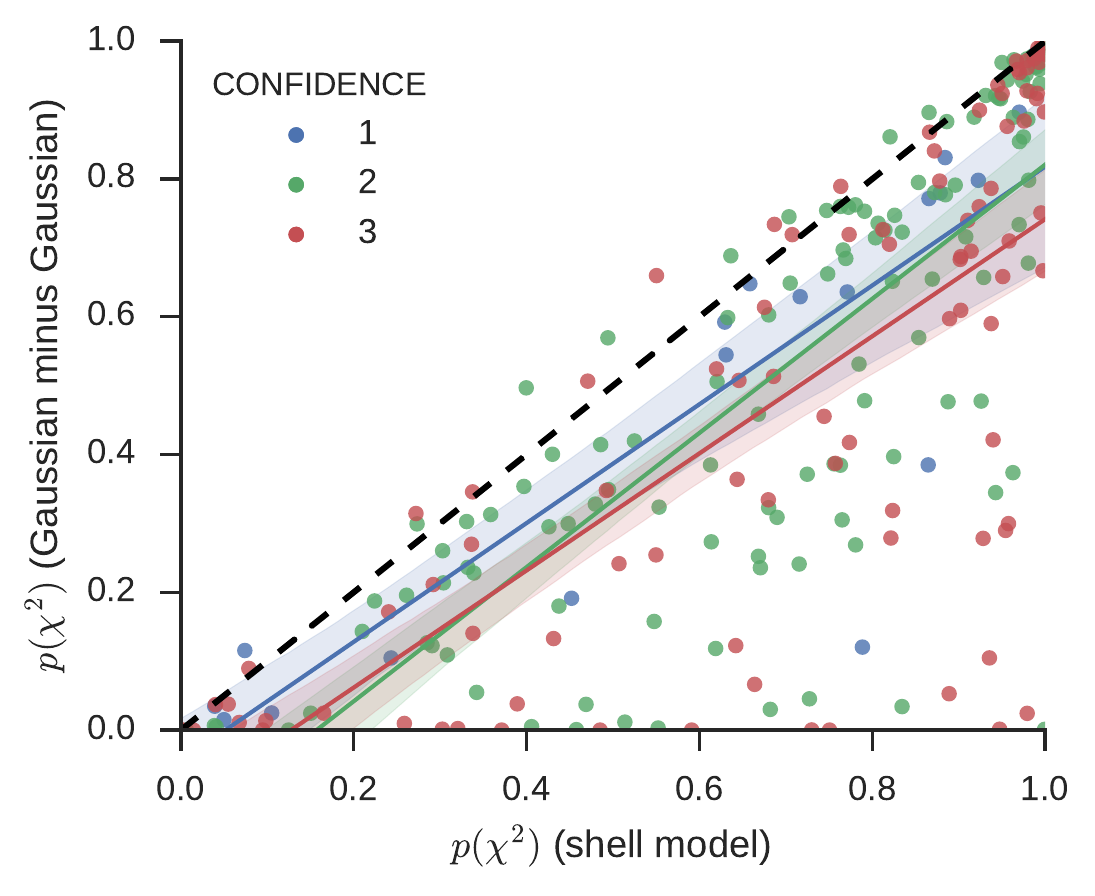}
  \caption{Quality of the fits of the `shell' versus the `Gaussian-minus-Gaussian' models. The color stands for the three `confidence' categories. Also included (as a `guide to the eye') are linear fits to the data, and the identity function (as black dashed line).}
  \label{fig:chisq_sm-vs-gmg}
\end{figure}

\begin{figure*}
  \centering
  \includegraphics[width=\textwidth]{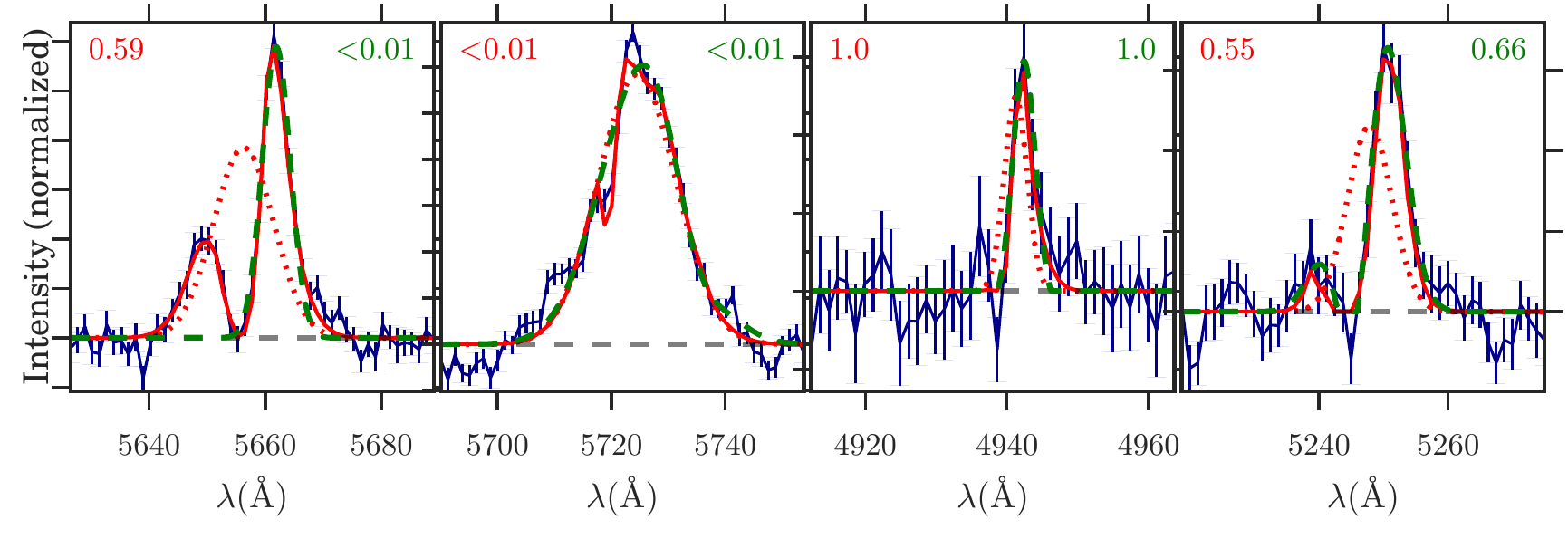}
  \caption{Example spectral fits of the `shell' and `Gaussian-minus-Gaussian' models with all combinations of fitting failures / successes. The blue lines show the data, the red solid line the shell model fit (with the intrinsic spectrum as dotted red line), and the green dashed line shows the `Gaussian-minus-Gaussian' best fit. The numbers in the panel show the $p(\chi^2)$ values of the best fits in the corresponding colors.}
  \label{fig:spectra_sm-vs-gmg}
\end{figure*}

\subsection{Contrasting the `shell-' with the `Gaussian-minus-Gaussian' model}
\label{sec:gmg}

As an alternative to the radiative-transfer based shell-model, we fit the $237$ \Lya spectra using the a `Gaussian-minus-Gaussian' model (which has been previously used in the literature modeling \Lya spectra; \citealp[e.g.,][]{2013MNRAS.428.1366J}). This model also consists of six free parameters, namely the location (i.e., the mean $\mu$) and width (the standard deviation $\sigma$) of two normal distributions which have been scaled by an arbitrary amplitude (the total area under the curve $A$). Please note that in concordance with the name, we do not allow negative amplitudes (which could correspond to a `Gaussian-plus-Gaussian' model), and also don't allow for negative fluxes, that is, we set the data point which would be negative to zero.

Figure~\ref{fig:chisq_sm-vs-gmg} shows a comparison between the fit qualities of the two models using the $p(\chi^2)$ values of the best-fitting parameters. The color coding stands again for the `confidence' category assigned to each spectrum. Clearly, the shell-model can reproduce most spectra better than the (even simpler -- but with the same number of free parameters) `Gaussian-minus-Gaussian' model leading to higher $p(\chi^2)$ values for $\sim 9$\% of the spectra\footnote{Overall, $83\%$ [$52\%$] of the `Gaussian-minus-Gaussian' fits possess a $p(\chi^2)>0.1$ [$>0.5$].}. In particular, in some cases the `Gaussian-minus-Gaussian' model fails to reproduce the observed spectral shape completely (i.e., $p(\chi^2) \lesssim 0.1$) while the shell-model can fit same spectrum successfully.
Overall, this improvement of the fit quality seems slightly larger for higher quality spectra (that is, higher `confidence' values); however, this trend is not statistically significant with the current data set.
On the other hand, spectra that cannot be modeled using the shell-model can neither be reproduced using the `Gaussian-minus-Gaussian' fit. Instead, both models yield comparable low $p(\chi^2)$ values.\\

Figure~\ref{fig:spectra_sm-vs-gmg} illustrates the reason for superiority of the `shell-' over the `Gaussian-minus-Gaussian'-model. Each panel shows an observed \Lya spectrum (in blue), the best shell-model fit (in solid red with dotted red the intrinsic spectrum), and (in dashed green) the best `Gaussian-minus-Gaussian' fit. The two numbers in the top left and right corner of each panel show the $p(\chi^2)$ values of the best-fit shell-model and `Gaussian-minus-Gaussian' fit, respectively. This means, the leftmost panel of Fig.~\ref{fig:spectra_sm-vs-gmg} shows an example where the shell-model yields a much better quality fit. The reason here is clearly the double peaked nature which cannot be captured by the `Gaussian-minus-Gaussian' model. The second panel from the left shows the same spectrum as in Fig.~\ref{fig:spectra_worstfit}, i.e., a case where the shell-model cannot reproduce the observed spectrum well. However, fitting two Gaussian curves yields a similar fit and, thus, also $p(\chi^2) < 0.01$. As mentioned in \S~\ref{sec:fit-quality} the reason for these low $p(\chi^2)$ values in this case are the small error bars on the data combined with the slight negative flux bluewards of the line. It is noteworthy that such wide, relatively symmetric spectra can naturally be well reproduced using a simple Gaussian curve. This means per construction the `Gaussian-minus-Gaussian' model suits this spectral type well. However, also the shell-model can feature such a spectrum -- given a low optical depth of the shell (through low column density and high outflow velocity) and a wide intrinsic spectrum. This will lead to hardly any radiative transfer effects, and thus, to a Gaussian emergent spectrum.

The third panel (from the left) in Fig.~\ref{fig:spectra_sm-vs-gmg} displays an single-peaked, asymmetric spectrum. This is an example where both models yield good fits to the data. This is also the case in the right panel of Fig.~\ref{fig:spectra_sm-vs-gmg} (which we chose because it is the spectrum where the $p(\chi^2)$ difference is greatest in favour for the `Gaussian-minus-Gaussian' model). Interestingly, here the two normal distributions can also reproduce the two peaks of the data. This is due to the lower blue peak (compared to the first panel) and the larger error within it.\\

A potential physical interpretation of the `Gaussian-minus-Gaussian' model is some intrinsic spectrum (e.g., shaped by within the interstellar medium) which is then processed by the circum- and / or intergalactic medium (see, e.g., \citealp{Gronke2016a} and \citealp{Dijkstra2007a}, respectively). The latter step would lead to a `\Lya halo' surrounding the galaxy and -- as (with increasing distance from the galaxy) most scatterings are out of the line-of-sight -- can be modeled as an effective absorption. This would mean that an increased level of absorption should correlated positively with the halo size.
However, we do not find any correlation of the `absorbed flux'\footnote{We calculate the absorbed flux ratio as $F_{\mathrm{abs}} = A_1 - \int f_{\mathrm{GMG}}(\lambda)\dd\lambda$ where $A_1$ is the amplitude of the positive Gaussian, and $f_{\mathrm{GMG}}(\lambda)$ is the best fit (normalized to unity).} ($r_{\mathrm{S}}\approx -0.07$) with $R_{\mathrm{Kron}}$. Neither we find any correlation with any other halo property discussed in \S~\ref{sec:haloprops}.
As both the intrinsic spectrum as well as the absorption feature may differ from a normal distribution -- which makes the `Gaussian-minus-Gaussian' model a simplification -- this result is maybe not too surprising.

One can surely extend the `Gaussian-minus-Gaussian' model with additional free parameters (by, e.g., allowing for negative amplitudes and non-zero skewness) in order to provide better fits to the observed \Lya spectra. However, one should keep in mind that the advantage of a radiative-transfer based model (such as the shell-model) is that its spectral shapes are at least a subset of all the possible \Lya spectra. This is not necessarily so for ``more artificial'' models. Also, these models to not directly alter the spectral shape by changing the frequency diffusion process, and, thus it physical content of the fitting parameters is even more questionable than in the shell-model.

\section{Conclusions}
\label{sec:conclusion}

We fitted the publicly available \Lya spectra of the ``MUSE-Wide survey'' published in \citet{Herenz2017} using the shell-model. This data-set contains $237$ spectra around redshift $z\sim 3.8$. We carried out the fitting procedure using the pipeline described in \citep{Gronke2015} which yields robustly the best-fit parameters as well as the respective confidence intervals. Our main findings can be described as follows:

\begin{itemize}[itemsep=10pt]
\item Overall, the shell-model provides excellent fits to a majority of the spectra (with $>90\%$ possessing a $p(\chi^2)>0.1$; see \S~\ref{sec:fit-quality}) with a better fits than the `Gaussian-minus-Gaussian' model (\S~\ref{sec:gmg}). Even the worst shell-model fits can reproduce the spectral shape reasonably well (cf. Fig.~\ref{fig:spectra_worstfit}). This implies that this simple model can practically fit all $237$ spectra.
  
\item We identified common parameter ranges which are typical for observed spectra. This allows future studies to generate more realistic \Lya spectra, and to compare newly obtained shell-model parameters to this distribution. For this purpose, we will make all the fits publicly available\footnote{\url{http://bit.ly/a-spectra-of-MW}}. With the exception of the shell's dust optical depth we find none of the parameters completely unphysical (see \S~\ref{sec:fitting-results}), i.e., within the range what has been derived by other means. For instance, we recovered from only the spectral shape outflows in the range of $\sim 200\kms$ (in particular, the higher quality spectra) which has been found using metal-absorption lines \citep[e.g.,][]{Steidel2010a}.
  
\item Two recovered parameter distributions \citep[which can have physical meaning in a multiphase medium with many clumps per sightline; see][]{2017arXiv170406278G} are of particular interest: \textit{(i)} the majority of spectra ($\sim 70\%$) are better fit through an outflowing shell and only $\sim 5 \%$ are prefer an inflowing shell. Given firmer underlying data this could yield insights into the gas budget of high-redshift galaxies. \textit{(ii)} half of the spectra can be fit with column densities of $N_{\HI} \lesssim 10^{17}\cm^{-2}$. This suggests a possible large fraction of Lyman-continuum leakers in the MUSE-Wide data set and supports the importance of \Lya emitting galaxies on the ionizing budget \citep{Dressler2014,Verhamme2014,Dijkstra2016b}.
  
\item We did not find any strong correlation neither amongst the shell-model parameters nor with the \Lya halo properties. However, a weak correlation between $\Delta v$ and $N_\HI$, and a tentative correlation between $\sigma_{\mathit{i}}$ and $\Delta v$ exist.
\end{itemize}

This work serves also as a `proof-of-concept', i.e., that the modeling of a large set of observed \Lya spectra -- as will be common in the near future -- is possible. A similar analysis can be improved by increasing the signal-to-noise ratio of the spectra (cf. Fig.~\ref{fig:sm_error}), and by providing priors on the systemic redshift (thus, confining $\Delta v$ to values close to zero).

\begin{acknowledgements}
  I thank the anonymous referee for constructive and helpful feedback.
  I am grateful to my PhD supervisor Mark Dijkstra for his guidance and comments on the draft.
  Furthermore, I thank Edmund Christian Herenz for a careful read of the draft. Also, I would like to thank him and the entire MUSE team for making the data used in this work publicly available. Finally, I wish to thank Lutz Wisotzki for inviting me to the Leibnitz Institute for Astrophysics in Potsdam.
  This research made use of a number of open source software such as the \texttt{Python} programming language and its packages \texttt{astropy} \citep{2013A&A...558A..33A}, \texttt{IPython} \citep{PER-GRA:2007}, \texttt{matplotlib} \citep{Hunter:2007}, \texttt{seaborn} \citep{seaborn}, and \texttt{SciPy} \citep{jones_scipy_2001}.
\end{acknowledgements}

\bibliography{references_all}

\end{document}